\documentclass[journal=jpclcd,manuscript=article,articletitle=true,layout=twocolumn]{achemso}

\usepackage{amsmath}
\usepackage{amssymb}
\usepackage{latexsym}
\usepackage{graphicx}
\usepackage{epsfig,epsf,rotating}
\usepackage{indentfirst}
\usepackage{epstopdf}
\usepackage{xcolor}
\usepackage{multirow}
\usepackage{tabularx}
\usepackage{ctable}

\usepackage{upgreek}
\newcolumntype{C}{>{\centering\arraybackslash}X}
\usepackage{nicefrac}

\title{Quantifying the Errors Introduced by Continuum Scattering Models on the Inferred Structural Properties of Proteins\footnote{
Notice:  This manuscript has been authored by UT-Battelle, LLC, under contract DE-AC05-00OR22725 with the US Department of Energy (DOE). The US government retains and the publisher, by accepting the article for publication, acknowledges that the US government retains a nonexclusive, paid-up, irrevocable, worldwide license to publish or reproduce the published form of this manuscript, or allow others to do so, for US government purposes. DOE will provide public access to these results of federally sponsored research in accordance with the DOE Public Access Plan (https://www.energy.gov/doe-public-access-plan}}

\author{Rohan S. Adhikari}
\affiliation{Department of Chemical and Biomolecular Engineering, Rice University, 6100 Main St., Houston, TX 77005, USA}

\author{Dilipkumar N. Asthagiri}
\affiliation{Oak Ridge National Laboratory, One Bethel Valley Road, Oak Ridge, TN 37830-6012}
\email{asthagiridn@ornl.gov}

\author{Walter G. Chapman}
\affiliation{Department of Chemical and Biomolecular Engineering, Rice University, 6100 Main St., Houston, TX 77005, USA}
\email{wgchap@rice.edu}

\begin{document}
	\begin{abstract}

    Atomistic force fields that are tuned to describe folded proteins predict overly compact structures for intrinsically disordered proteins (IDPs). To correct this, improvements in force fields to better model IDPs are usually paired with scattering models for validation against experiments. For scattering calculations, protein configurations from all-atom simulations are used within the continuum-solvent model CRYSOL for comparison with experiments. To check this approach, we develop an equation to evaluate the radius of gyration (R$\mathrm{_g}$) for any defined inner-hydration shell thickness given all-atom simulation data. R$\mathrm{_g}$ based on an explicit description of hydration waters compares well with  the reference value of R$\mathrm{_g}$ obtained using Guinier analysis of the all-atom scattering model. However, these internally consistent estimates disagree with R$\mathrm{_g}$ from CRYSOL for the same definition of the inner-shell. CRYSOL can over-predict R$\mathrm{_g}$ by up to 2.5 $\mathrm{\AA}$. We rationalize the reason for this behavior and highlight the consequences for force field design. 
   \begin{tocentry} 
			\includegraphics[width = 1.00\textwidth]{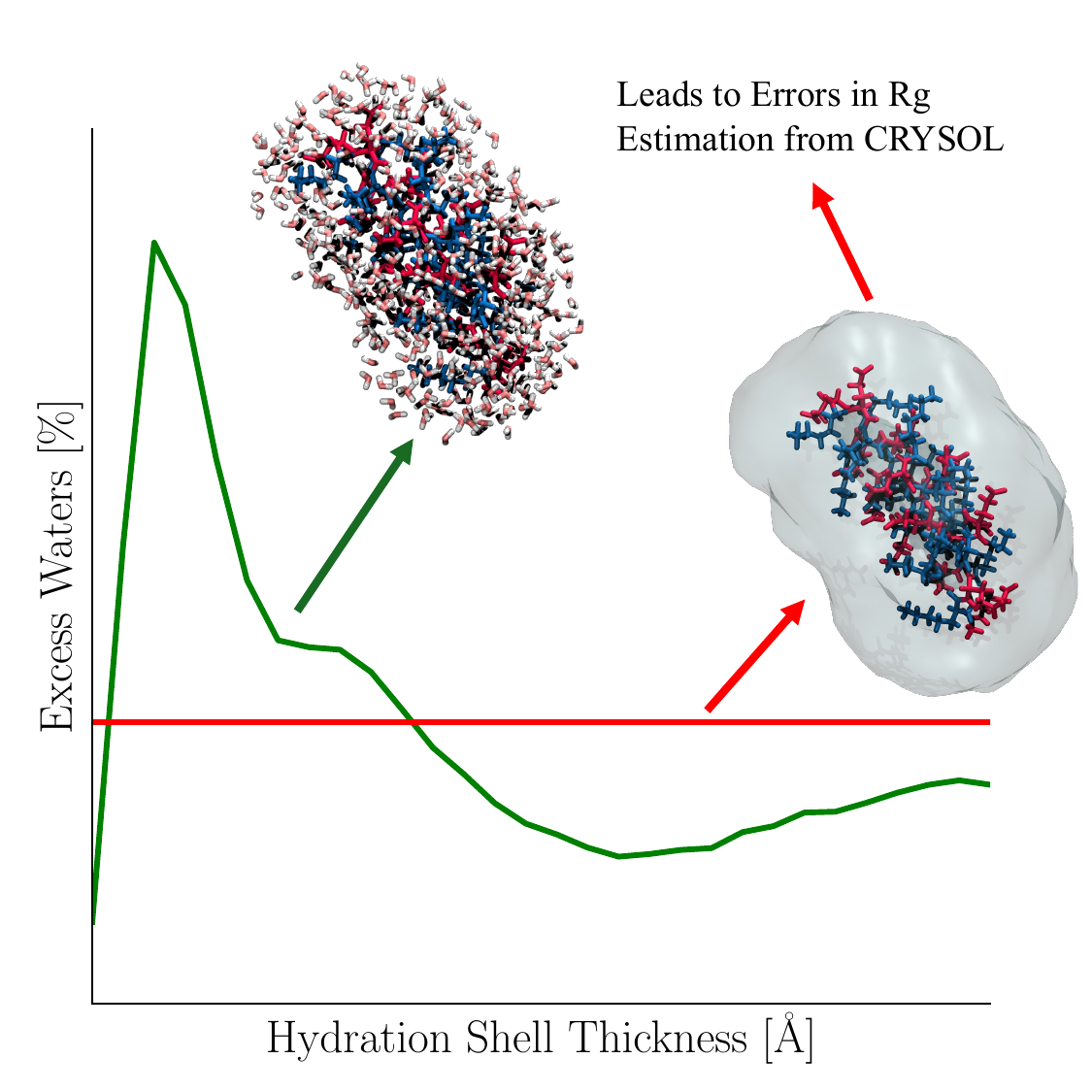} 
		\end{tocentry}

	\end{abstract}
	
	\maketitle
	
	Intrinsically Disordered Proteins (IDPs) have garnered tremendous scientific interest due to their ability to remain functional despite being disordered\cite{wright1999intrinsically, uversky2002natively, heery1997signature, janknecht1996transcriptional, romero1998thousands}. The consensus in the scientific community links their malleable structural properties to a variety of intracellular phenomena such as the formation of membraneless organelles\cite{wei2017phase, shin2017liquid} and the regulation of signalling networks\cite{bondos2022intrinsically, babu2011intrinsically}. When they misfire due to aggregation or mutation they are known to cause a host of pathologies such as amyloidosis\cite{uversky2008amyloidogenesis, mukhopadhyay2020dynamism} or cardiovascular diseases\cite{cheng2006abundance, monti2022amazing}. Traditional force field based simulation models were tuned to replicate the properties of folded proteins and have been shown to predict overly compact structures for IDPs \cite{rieloff2021molecular, henriques2015molecular, best2017computational, huang2018force, nerenberg2018new}. Numerous modifications to force fields have been proposed to circumvent this issue\cite{wang2014new, liu2019extensive, huang2017charmm36m, piana2015water, shabane2019general}. Such modifications need to be validated against experimental data. Due to the availability of high flux synchotron sources and data analysis programs, Small Angle X-ray Scattering (SAXS) is one of the most widely used methods in such studies\cite{lipfert2007small, bernado2012structural, svergun2013small}.    
	
 In SAXS both the solute and the solvent contribute to the scattered intensity.  Relative to a neat solvent,  
 the presence of the solute affects scattering by the solvent in two specific ways: (1) by excluding a volume that would otherwise be occupied by the solvent, and (2) by associating with an excess of solvent in the inner-hydration shell 
\cite{serdyuk2007methods}.  In the so-called process of background subtraction\cite{bernado2007structural, svergun2013small} the role of the bulk solvent is sought to be removed from the scattering profile of the protein-solvent mixture. The remaining signal is interpreted to infer properties of the protein conformation. It is clear that the specific role of the solute on the solvent needs to be treated as a part of the solute, and to this end, scattering models are necessary to  translate the simulated coordinates to experimentally measurable quantities. 

CRYSOL was the first model to account for the specific effects of the solute on the solvent. It does so by assuming a continuum bulk density of the solvent in the excluded volume region and a continuum excess density of the solvent in the solvation layer\cite{svergun1995crysol, manalastas2021atsas}. The program FoXS expedites the scattering computations by strategically placing dummy solvent atoms around the exposed surface of the solute in  solution\cite{schneidman2013accurate, schneidman2016foxs}. The importance of accounting for more density variation in the solvation layer was recognized by Park et.~al.~\cite{park2009simulated} who proposed an explicit water scattering model that incorporates the solvation layer in atomic detail. Virtanen et.~al.~used the detailed hydration layer predicted by the program HyPred\cite{virtanen2010modeling} to calculate the SAXS profiles of proteins\cite{virtanen2011modeling}. Similarly, Poitevin et.~al.~used the solvent maps generated by AquaSol\cite{koehl2010aquasol} in their program AquaSAXS\cite{poitevin2011aquasaxs}. All of the scattering models discussed so far assume a frozen solute which is enough to accurately describe the scattering in the Guinier regime. For a full SAXS comparison the thermal fluctuations of the solute also need to be included\cite{chen2014validating}. The program WAXSIS\cite{knight2015waxsis}, incorporating both the thermal fluctuation of the solute and an atomic solvation layer, is currently the gold standard for simulated X-ray scattering\cite{henriques2018calculation, hub2018interpreting}.  
 
Simulation studies have largely been indifferent to the improvements in scattering models noted above. CRYSOL is frequently the model of choice when force fields are validated against SAXS data \cite{rauscher2015structural, henriques2015molecular, rieloff2021molecular}. Most notably CRYSOL was used in the development of the CHARMM36m force field for IDP simulations \cite{huang2017charmm36m}. CRYSOL's computational speed and ease of use are two of the major reasons for its continued use despite the availability of more detailed models \cite{svergun2013small, manalastas2021atsas}. Even though the assumption of a uniform excess density in the hydration layer has been called into question by several explicit water simulation studies\cite{chen2014validating, virtanen2011modeling, linse2023scrutinizing}, the consequences of this approximation on the scattering predictions made by CRYSOL have not been adequately understood. We believe it is imperative to define metric(s) that allow for the validation of scattering models by studying their predictions in isolation. We develop such a metric. 
 
    
    All of the scattering models in literature seek to mimic the experimental background subtraction procedure represented in Eq.~\ref{eq:back_sub}\cite{svergun2013small, grishaev2010improved}.
    \begin{eqnarray}
     \Delta I(q) = I_{\mathrm{A}}(q) - I_{\mathrm{B}}(q)
    \label{eq:back_sub}
    \end{eqnarray}

\begin{figure*}[ht!]
		\begin{center}
			\includegraphics[width=1.0\textwidth]{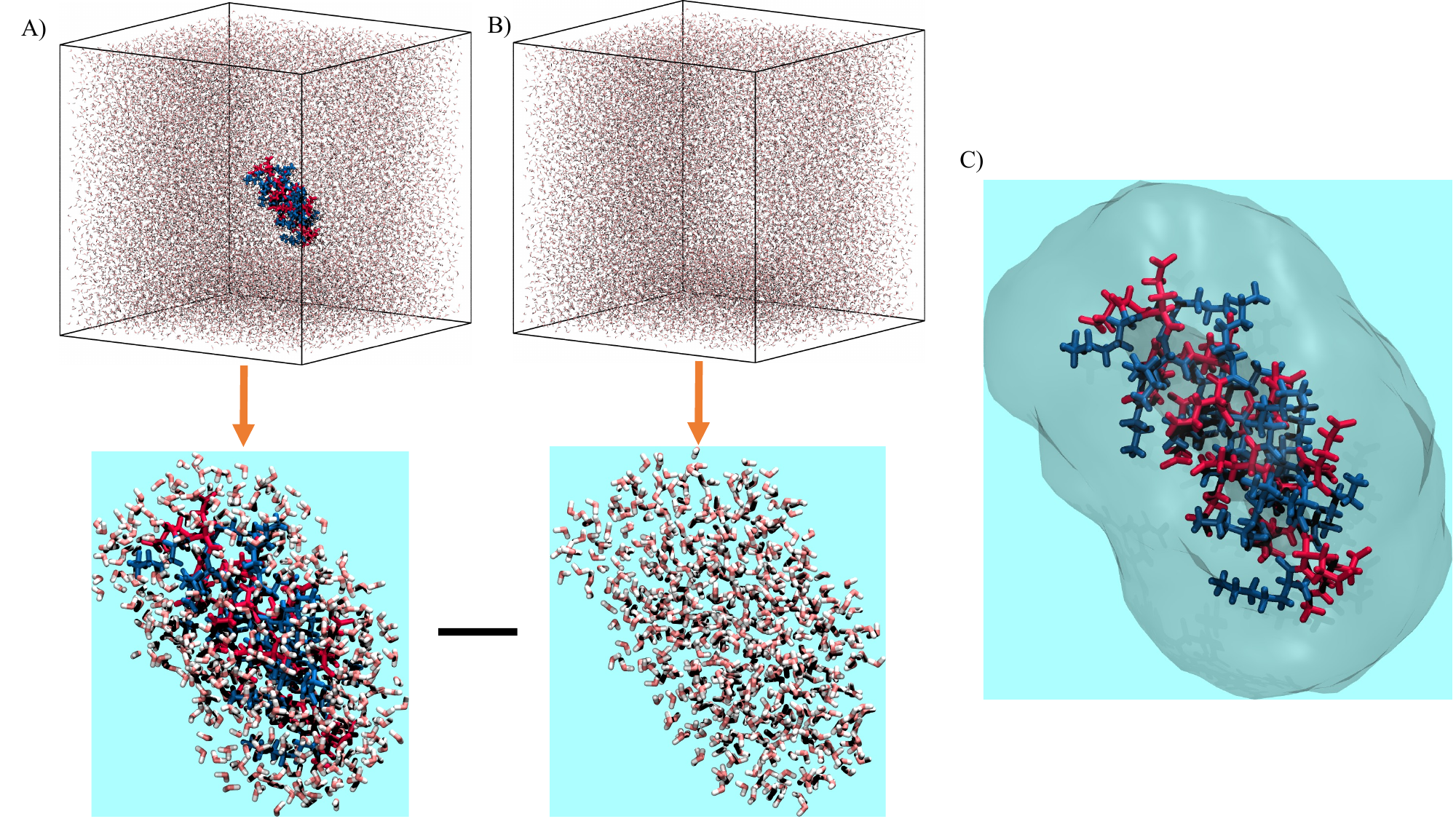}
		    \caption{Schematic representation of scattering models. Panels (A) and (B) together constitute the explicit water scattering model. The intensity from a bulk water simulation (B) is subtracted from the intensity of a protein-water simulation (A). The continuum scattering model is represented in (C). The blue background in (A), (B) and (C) represents the bulk-like solvent which is treated as a continuum.} \label{fig:sca_models}
		\end{center}
	\end{figure*}

    Park et.~al.\cite{park2009simulated} came up with an equation that computes the background subtracted intensities for simulated structures of biomolecules. One of the advantages of their formalism is the ability to incorporate an atomistic hydration layer. The simulated background subtraction procedure is graphically represented in Fig.~\ref{fig:sca_models}A and \ref{fig:sca_models}B. The framework developed in Ref.~\citenum{park2009simulated} which we shall henceforth call the explicit water scattering model is shown in Eq. \ref{eq:expl_sca}.
      \begin{align}
        \Delta I(q) = & \bigg \langle \big | \langle A(\textbf{q}) \rangle - \langle B(\textbf{q}) \rangle \big | ^2 + \nonumber \\ 
        & \big[ \langle |A(\textbf{q})|^2 \rangle  - \big | \langle A(\textbf{q}) \rangle \big |^2 \big] - \nonumber \\ 
        & \big[ \langle |B(\textbf{q})|^2 \rangle  - \big | \langle B(\textbf{q}) \rangle \big |^2 \big] \bigg \rangle_{\mathrm{\Omega}}
        \label{eq:expl_sca}
    \end{align}   
    $A$ and $B$ in Eq.~\ref{eq:expl_sca} correspond to the amplitudes from the protein-water simulation and the bulk water simulation, respectively. An envelope of a defined volume is carved out of the protein-water simulation box (Fig.~\ref{fig:sca_models}A). A similar envelope of the same volume is carved out of the bulk water simulation box (Fig.~\ref{fig:sca_models}B). The behavior of waters outside this envelope is assumed to be bulk-like. The volume of the envelope is usually defined based on the distance from solute atoms. Moving a distance 7 $\mathrm{\AA}$ away from all solute atoms is known to be enough to reach bulk-like behavior of water \cite{park2009simulated, chen2014validating}. In our work the hydration shell is defined to be 3 $\mathrm{\AA}$ thick from the surface of the solute atoms to maintain consistency with the CRYSOL description \cite{krywka2007small, svergun2013small}. This choice also ensures that the binding energy of the solute with the solvent outside the inner-shell admits a Gaussian description 
    \cite{tomar:jpcb16,app:jpcb21,adhikari2022hydration}, signalling non-specific interactions. 

    The CRYSOL model (Eq.~\ref{eq:crysol})
    \begin{eqnarray}
        \Delta I(q) = \langle | A_{\mathrm{p}} (\textbf{q}) - \rho_{\mathrm{b}}A_\mathrm{ex}(\textbf{q}) + \delta \rho_\mathrm{h} A_\mathrm{h} (\textbf{q}) |^2 \rangle_{\mathrm{\Omega}}  
    \label{eq:crysol}
    \end{eqnarray}
     is a simplification of the explicit water scattering model. In a continuum density description, the terms in Eq.~\ref{eq:expl_sca} which account for density fluctuations vanish\cite{park2009simulated}, 
     and the amplitude difference between the protein-water system and the bulk water system is defined in terms of an excess density parameter ($\delta \rho_h$) for the hydration shell and a volume parameter ($V_{ex}$) for the excluded volume\cite{svergun1995crysol}.

    The CRYSOL model can be used either by inputting a hydration layer contrast or by varying the parameters in Eq.~\ref{eq:crysol} to get the best fit to the experimental data\cite{manalastas2021atsas}. The CRYSOL model does not require a bulk water simulation to perform the subtraction and is graphically represented in Fig.~\ref{fig:sca_models}C.


Our internal consistency check to quantify the errors that arise from a continuum hydration description (Eq.~\ref{eq:crysol}) is as follows.  For a defined inner-hydration shell thickness, Eq.~\ref{eq:coor_rg} is used to calculate the R$\mathrm{_g}$ from the coordinates of the protein-hydration layer complex. 
($N_A$ and $N_B$ are the total number of atoms in the protein and the hydration layer and the number of atoms in the same volume of a bulk water simulation respectively. $N_F$ is the total number of simulated frames. See the SI for further details.)
\begin{align}
        &R_{g}^{coor} = \label{eq:coor_rg}\\ 
        &\sqrt{\frac{\sum\limits_{N_F} \sum\limits_{N_A} n_{A}^{e} |\mathbf{r_{A}} - \mathbf{r_{com}}|^2 - \sum\limits_{N_F} \sum\limits_{N_B} n_{B}^{e}|\mathbf{r_{B}} - \mathbf{r_{com}}|^2}{\sum\limits_{N_F} \sum\limits_{N_A} n_{A}^{e} - \sum\limits_{N_F} \sum\limits_{N_B} n_{B}^{e}}}, \nonumber
    \end{align}
    where, 
    \begin{eqnarray}
         \mathbf{r_{com}} = \frac{\sum\limits_{N_F} \sum\limits_{N_A} n_{A}^{e} \mathbf{r_{A}} - \sum\limits_{N_F} \sum\limits_{N_B} n_{B}^{e}\mathbf{r_{B}}}{\sum\limits_{N_F} \sum\limits_{N_A} n_{A}^{e} - \sum\limits_{N_F} \sum\limits_{N_B} n_{B}^{e}}
    \label{eq:coor_com}
    \end{eqnarray}
    
 The R$\mathrm{_g}$ calculated using Eq.~\ref{eq:coor_rg} is
 compared to the R$_{\mathrm{g}}$ predicted from Guinier analysis\cite{cantor1980biophysical, zheng2018extended} of the scattering curve. (Fig.~S1 provides a schematic of the consistency check.) In our work the hydration layer definition in Eq.~\ref{eq:coor_rg} is consistent with CRYSOL and hence the R$_{\mathrm{g}}$ from the two methods can be directly compared. Inconsistency between these measures serves to signal potential
 failures of either the definition of the inner-shell thickness or the continuum approximation or both. 
 
 When MD simulations are paired with CRYSOL and validated against experimental data the errors from the scattering model alone cannot be isolated. By assuming a hydration shell that is 3 $\mathrm{\AA}$ thick and by supplying the contrast calculated from explicit water simulations (see the SI for details) as an input to CRYSOL, we ensure that there is a direct correspondence between the explicit water scattering model (Eq.~\ref{eq:expl_sca}), the R$\mathrm{_g}$ calculation from coordinates (Eq.~\ref{eq:coor_rg}) and the CRYSOL model (Eq.~\ref{eq:crysol}). In this way we have devised a consistency check that can isolate the errors due to a continuum description of the hydration layer. 

\begin{figure*}[ht!]
		\begin{center}
            \vspace{-1.8cm}
			\includegraphics[width=1.0\textwidth]{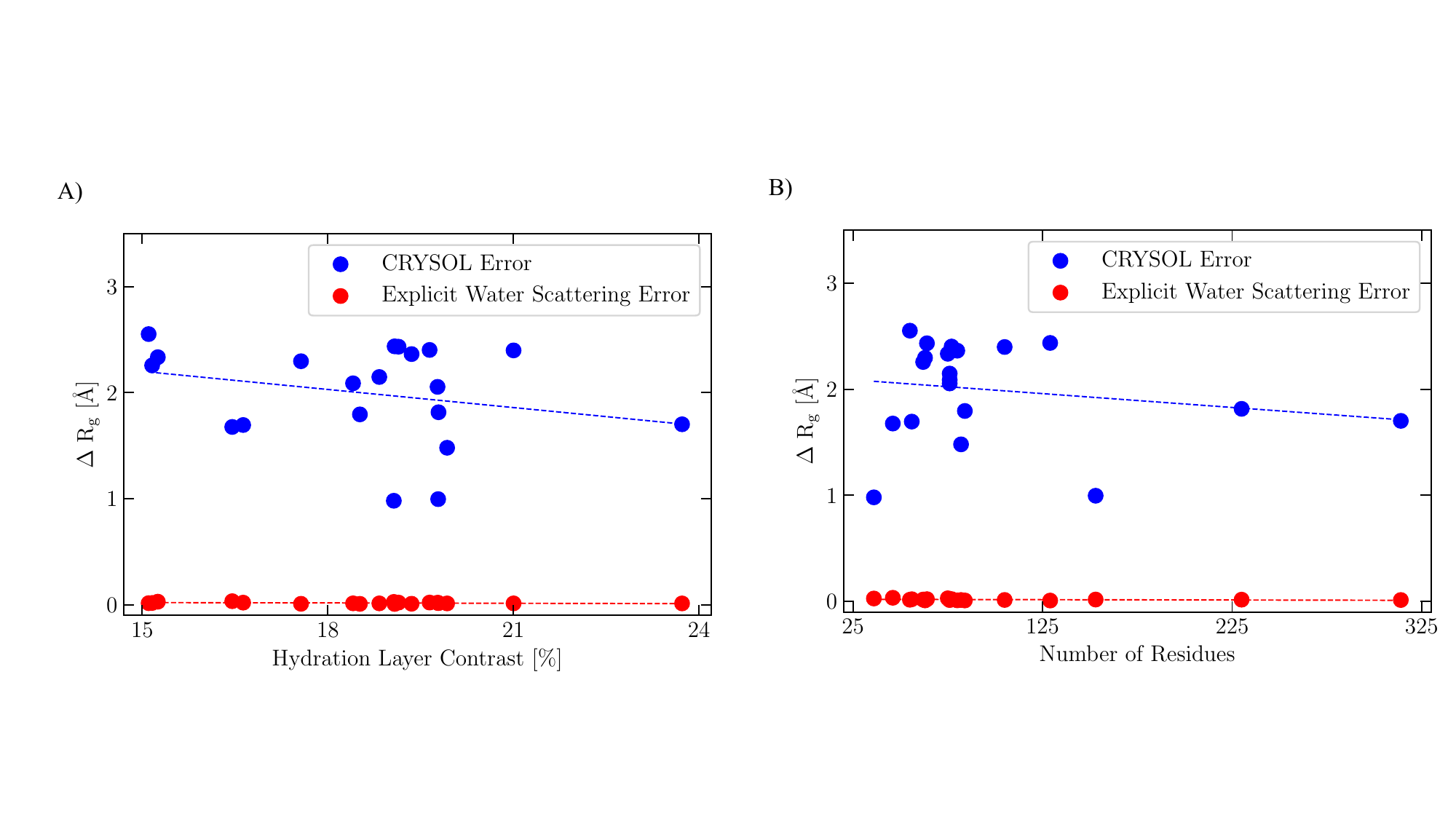}
            \vspace{-2.2cm}
		    \caption{Errors from the explicit water scattering model are shown in red. Errors from CRYSOL are shown in blue. The errors from the scattering models are plotted against the hydration layer contrast in (A) and the size of the protein in (B). Best fit lines are a guide to the eye.} \label{fig:results_1}
		\end{center}
	\end{figure*}

We chose 20 protein structures from the Protein Data Bank (PDB). The same 20 proteins were used to predict the hydration layer around proteins in the HyPred\cite{virtanen2011modeling} program. The proteins were simulated in explicit water using
NAMD\cite{phillips2020scalable}. The solute was frozen during the simulation and was modeled using the CHARMM36m (C36m)\cite{huang2017charmm36m} force field. Modified TIP3P waters (m-TIP3P)\cite{neria1996simulation} were used in all simulations. All of the explicit water scattering calculations and the R$_{\mathrm{g}}$ calculations were performed using in-house PYTHON scripts. The MDTraj\cite{McGibbon2015MDTraj} library was used to load the NAMD trajectory files. All of the CRYSOL calculations were performed using CRYSOL 3.0\cite{manalastas2021atsas}. Fig.~\ref{fig:results_1} shows the errors from CRYSOL and the explicit water scattering model.

 $\Delta$R$_{\mathrm{g}}$ in Fig.~\ref{fig:results_1} represents the difference between the R$_{\mathrm{g}}$ prediction made by the scattering model and the R$_{\mathrm{g}}$ calculated from the coordinates (Eq.~\ref{eq:coor_rg}). $\Delta$R$_{\mathrm{g}}$ for CRYSOL is always positive and is as high as 2.5 $\mathrm{\AA}$. The errors from CRYSOL remain significant irrespective of the contrast in the hydration layer (Fig.~\ref{fig:results_1}A) or the size of the protein (Fig.~\ref{fig:results_1}B). $\Delta$R$_{\mathrm{g}}$ for the explicit water scattering model is very close to  $0\, \mathrm{\AA}$. Hence it represents the information in the system accurately.

 CRYSOL can also be used to fit to the experimental data by varying the parameters in Eq.~\ref{eq:crysol}. The problem of overfitting in CRYSOL has previously been studied for protein-detergent complexes\cite{chen2015structural}. However quantifying the errors in R$_{\mathrm{g}}$ from this approach is not easy, because the experimental data does not correspond to the solute alone but is a property of the protein-hydration layer complex. A variety of solute structures with different R$_{\mathrm{g}}$ can be combined with appropriate excess density parameters in CRYSOL to obtain good fits against the experimental data. The fitting procedure used in CRYSOL is represented by Eq.~\ref{eq:chi2}
  \begin{eqnarray}
        \chi^2 = \frac{1}{N}\sum_{i=1}^{N} \bigg[\frac{fI_{calc}(q_i) - I_{ref}(q_i)}{\sigma_{ref}(q_i)} \bigg]^2,
    \label{eq:chi2}
    \end{eqnarray}

 where $I_{calc}(q_i)$ are the intensities predicted by CRYSOL, $I_{ref}(q_i)$ are the intensities predicted by the explicit water scattering model, and $\sigma_{ref}(q_i)$ are the error bars from the explicit water scattering model. A factor of $f$ (Eq.~\ref{eq:chi2}) is used in the fitting procedure to scale the predictions from the scattering model to the same units as the reference. This has no consequence to the shape of the scattering curve and hence the properties of interest. 

 Fitting to the experimental data does not provide any information about the difference in the solute R$_{\mathrm{g}}$ between experiments and CRYSOL. To circumvent this issue, we fit CRYSOL to the explicit water scattering data. This way the R$_{\mathrm{g}}$ of the underlying solute is known exactly in addition to the scattering data of the protein-water complex. The explicit water scattering data for reference structures of four different proteins are obtained by freezing the solute. Eight protein structures for each of the four proteins with R$_{\mathrm{g}}$ ranging from 2.5 $\mathrm{\AA}$ lower than the reference and 1 $\mathrm{\AA}$ higher than the reference are selected. The protein structure and the explicit water scattering data are input into CRYSOL. This workflow is shown in schematic Fig.~S2. 
 
\begin{figure*}[ht!]
		\begin{center}
			\includegraphics[width=1.0\textwidth]{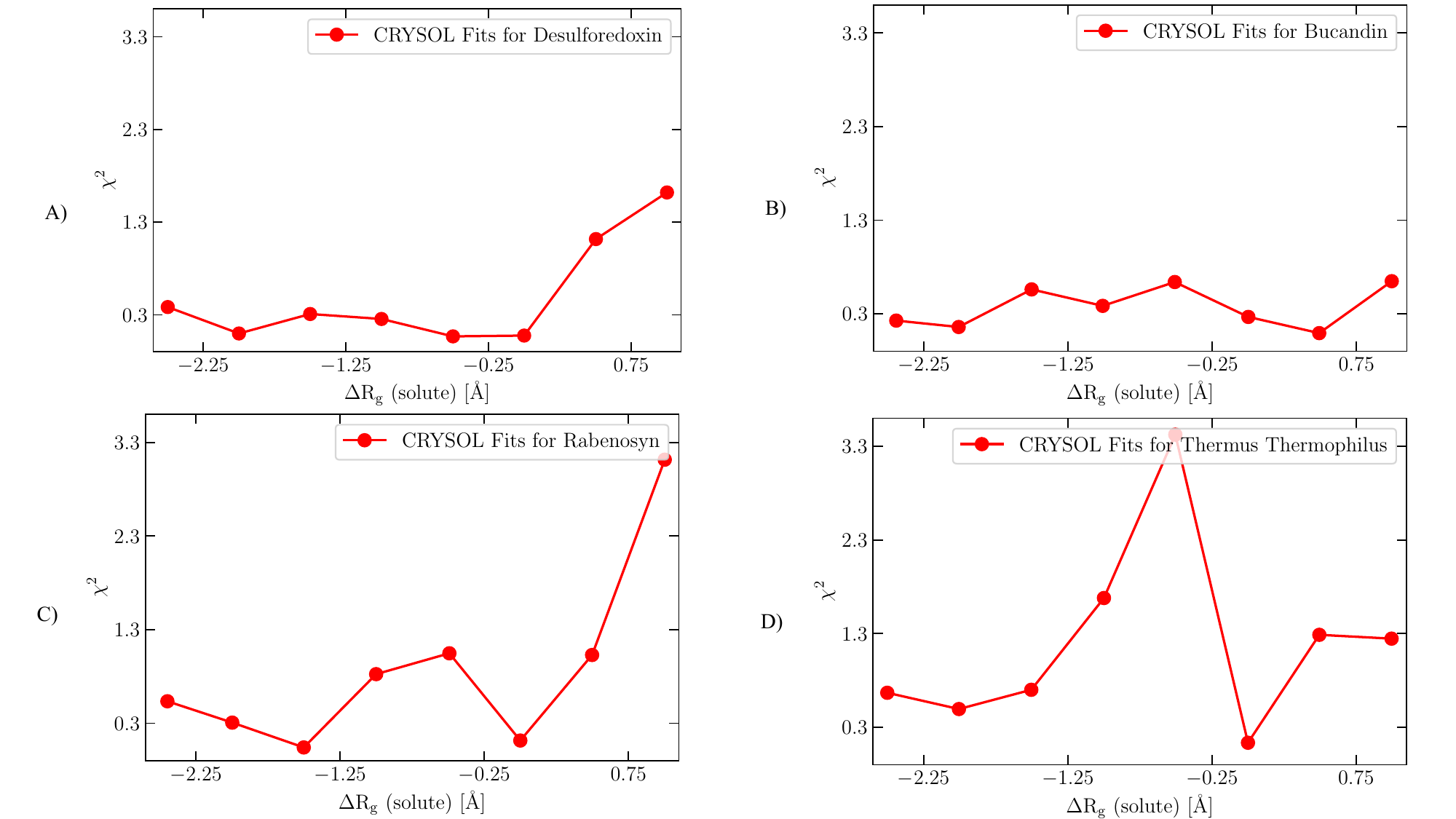}
            \vspace{-0.3cm}
		    \caption{$\mathrm{\chi^2}$ values for 8 structures of the protein plotted against the difference in R$_{\mathrm{g}}$ between the structure in question and the reference ($\mathrm{\Delta}$R$_{\mathrm{g}}$). The four proteins used in the study are (A) Desulforedoxin (PDB code: 1DXG), (B) Bucandin (PDB code: 1F94), (C) Rabenosyn (PDB code: 1YZM), and (D) Thermus Thermophilus (PDB code: 2ZQE).} \label{fig:chi2fitcrysol}
		\end{center}
\end{figure*}

 In Fig.~\ref{fig:chi2fitcrysol}, $\mathrm{\chi^2}$ values for all eight structures are plotted against the $\mathrm{\Delta}$R$_{\mathrm{g}}$ of the solute. $\mathrm{\Delta}$R$_{\mathrm{g}}$ is the difference between the R$_{\mathrm{g}}$ of the structure and the R$_{\mathrm{g}}$ of the reference. A $\mathrm{\chi^2}$ value less than 0.5 is considered an excellent fit (probability of the fit: 1.000), while a $\mathrm{\chi^2}$ value less than 1.0 is considered a good fit (probability of the fit: 0.460).
Using the $\chi^2$ measure, structures  $2.5\, \mathrm{\AA}$ \textbf{more compact} than the reference are frequently shown to be excellent fits by CRYSOL. Our results call into question this practice of using $\mathrm{\chi^2}$ as a measure of the accuracy of scattering models. CRYSOL does a better job of discriminating between structures that are more expanded than the reference. This is because of the innate proclivity of CRYSOL to overpredict the R$_{\mathrm{g}}$ of the system. Both by contrast matching and by fitting CRYSOL to the reference we find that it has a tendency to over-estimate the effect of the excess waters on the R$_{\mathrm{g}}$ of the system. We will explain the rationale for this divergence below.

\begin{figure*}[ht!]
		\begin{center}
            \vspace{-1.5cm}
			\includegraphics[width=1.0\textwidth]{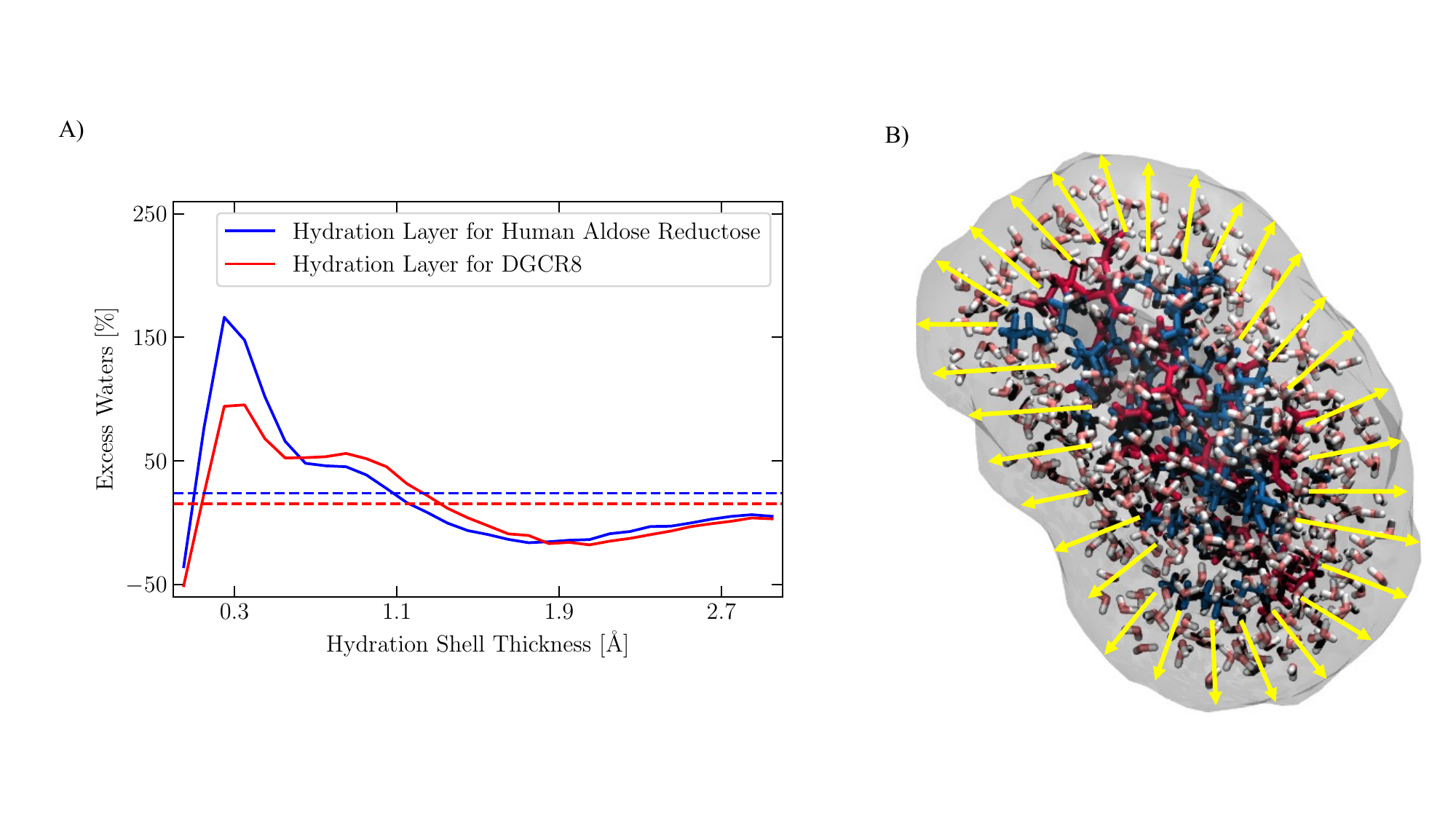}
            \vspace{-1.5cm}
		    \caption{The distribution of waters from an explicit water simulation (solid lines) are shown for two proteins as a function of the hydration shell thickness in (A). The two proteins studied are Human Aldose Reductose (PDB code: 1US0) and DGCR8 (PDB code: 3LE4). The continuum distribution for both proteins are shown in dotted lines in (A). The movement of the waters away from the protein surface which leads to the over-prediction by CRYSOL is graphically represented in (B).} \label{fig:discussion}
		\end{center}
	\end{figure*}

 To understand the overprediction from CRYSOL, let us look closely at the distribution of waters in the hydration layer. The distribution of waters for two proteins from explicit water simulations are shown in Figure \ref{fig:discussion}A. The distribution shows a characteristic sharp peak close to the surface of the protein (within 1 $\mathrm{\AA}$ from the surface) and then a dip in the density from about 1-3 $\mathrm{\AA}$ from the protein surface. Figure \ref{fig:discussion}A also shows the distribution of waters if it were a continuum (dotted lines). 

 In the continuum density framework, waters will have to be moved from the surface of the protein to the outer edges of the hydration shell to maintain a constant contrast. This movement of waters away from the protein surface (shown in Fig.~\ref{fig:discussion}B) increases the R$_{\mathrm{g}}$ of the protein-water complex and significantly distorts the information in the system. Henriques et.~al.\cite{henriques2018calculation} reported that good fits were obtained by CRYSOL only when the contrast was set to unphysically low values (around 3 \%). The contrast for proteins is expected to be in the range of 5-20 \%\cite{svergun2013small}. The findings in Ref.~\citenum{henriques2018calculation} further enunciate our point about the over-prediction in R$_{\mathrm{g}}$ from CRYSOL. The alternate approach to using a low contrast within the continuum description would be to vary the hydration shell thickness. However, the full consequences of these approaches to the underlying physics of scattering need to be understood first. 

Thermal fluctuations of the protein are not considered in this work for which there are two main reasons. Firstly, the continuum scattering equation used in CRYSOL is not easily extendable to account for thermal fluctuations. Some groups have attempted to use CRYSOL to understand the physics of fluctuating proteins but the ensemble averaging techniques used in these studies are questionable and need further investigation\cite{rauscher2015structural, yang2009rapid}. Second and most importantly, mere fluctuations alone do not cause a deviation in R$_{\mathrm{g}}$ which is the primary quantity of interest here. The arguments and data for this claim are provided in the Sec.~S6 (SI).


 In addition to the drawbacks of the continuum density description (addressed here) there are also questions about the definition of the hydration shell in CRYSOL. CRYSOL uses an angular envelope function to geometrically define the shape of the protein and its surrounding hydration shell\cite{svergun2013small, svergun1995crysol}. This method is known to be accurate for relatively simple shapes (globular proteins) but its applicability to the conformations of IDPs in solution is questionable and needs to be investigated further\cite{svergun2013small}. The division of the envelope into an excluded volume and the hydration shell is a requirement of the continuum solvent approach. Since the explicit water scattering model uses an atomistic description of the solvent such a division is not necessary. This further highlights the need for using the explicit water scattering model in the validation of IDP-specific force field improvements.

 On a larger note, the equation developed in this work to calculate the R$_{\mathrm{g}}$ directly from the coordinates of an explicit water simulation (Eq.~\ref{eq:coor_rg}) provides a credible consistency check to validate the Guinier analysis based R$_{\mathrm{g}}$ prediction from any scattering model. R$_{\mathrm{g}}$ is often the quantity used in simulation studies to estimate the ability of force field improvements to capture the properties of extended IDP ensembles\cite{shabane2019general, huang2017charmm36m, rauscher2015structural}. When understood in this context the importance of the methodology developed in this work becomes clearer. Zheng et.~al.\cite{zheng2018extended}~have suggested that the q range (usually a q$\mathrm{_{max}}$ of 1.3/R$_{\mathrm{g}}$) used in the Guinier analysis of experimental scattering profiles is too broad and can lead to errors in R$_{\mathrm{g}}$ estimation by up to 10\% for IDPs. A q$\mathrm{_{max}}$ of 1.3/R$_{\mathrm{g}}$ is used in experiments due to the difficulty in obtaining high quality background subtracted scattering data at lower q values. However, this problem does not arise for theoretical scattering models. The q$\mathrm{_{max}}$ value in theoretical models can be set as close to zero as required to conform accurately to the approximations in Guinier analysis. 
 
 The explicit water scattering model displays a high degree of consistency to the data obtained from simulations. The R$_{\mathrm{g}}$ of the solute alone is also known in addition to the scattering curve which is not possible in experiments. These advantages were used here to show that it is possible to obtain good $\mathrm{\chi^2}$ fits from CRYSOL even for solute structures that are significantly more collapsed than the reference. The methodology used in this work for CRYSOL can be applied to quantify the over-fitting from any general scattering model with free parameters.

 Several methods of generating simulated structures of IDPs have been proposed recently and are in need of validation. In our work so far we have explained the limitations of popular implicit solvent models in predicting accurate conformations for polyampholytes\cite{shi2023influence, adhikari2022hydration}. In the near future we also plan to evaluate the performance of IDP-specific improvements to explicit water force fields. Scattering models when used correctly can act as powerful discriminatory tools that aid in the further development of such models.

 In conclusion, we have developed a methodology here that allows for an effective validation of scattering models. By isolating the predictions from the scattering model, the methodology negates any possible bias in the analysis that could arise from external errors (such as errors from force fields). Our results suggest that the commonly used $\mathrm{\chi^2}$ metric is not a reliable indicator of the accuracy of scattering models. By matching the hydration layer contrast from explicit water simulations, we found CRYSOL to distort the R$_{\mathrm{g}}$ of the system by about 2 $\mathrm{\AA}$. By fitting CRYSOL to the explicit water scattering data, its tendency to over-predict the R$_{\mathrm{g}}$ of the system was re-established from a different perspective.
 This is particularly significant since the difference in the R$_{\mathrm{g}}$ predictions from contemporary force fields is on the order of 2-3 $\mathrm{\AA}$. The errors from the explicit water scattering model were almost 2 orders of magnitude lower than those from CRYSOL. The average absolute error from the explicit water scattering model was about 0.02 $\mathrm{\AA}$ (1.99 $\mathrm{\AA}$ from CRYSOL), which is of no consequence to the studies of interest. Hence we strongly recommend the use of the explicit water scattering model in foundational studies such as force field validation.

\section{Supporting Information}
{The Supporting Information includes (1) a schematic of the consistency check and $\chi^2$ over-fitting, (2) additional simulation details, (3) details on the contrast and excluded volume calculation, (4) details on scattering computation and R$_{\mathrm{g}}$ estimation, (5) scattering code validation, and (6) a note on the effect of the thermal fluctuations on the R$_{\mathrm{g}}$ of the system.}

\section{Acknowledgements}
	We thank Dr.~Amanda B.~Marciel and Winnie H.~Shi for engaging discussions about SAXS. RSA would like to thank Dr.~Jochen S.~Hub for his insights on the theory of scattering models.
 DA thanks Fred Poitevin for encouraging comments. We thank Dr.~Arjun Valiya Parambathu for his help in creating images. 
	We thank the Robert A.~Welch foundation for financial support (Grant No.~C-1241). We gratefully acknowledge the computational time on Anvil GPUs at the Rosen Center for Advanced Computing (RCAC), Purdue University, and on Stampede2 at the Texas Advanced Computing Center (TACC), University of Texas at Austin. Computational time on these clusters was obtained under a National Science Foundation (NSF) Advanced Cyberinfrastructure Coordination Ecosystem (ACCESS) project (Allocation no.~CHM-230024).
	Research at Oak Ridge National Laboratory is supported under contract DE-AC05-00OR22725 from the U.S. Department of Energy to UT-Battelle, LLC.


\providecommand{\latin}[1]{#1}
\makeatletter
\providecommand{\doi}
  {\begingroup\let\do\@makeother\dospecials
  \catcode`\{=1 \catcode`\}=2 \doi@aux}
\providecommand{\doi@aux}[1]{\endgroup\texttt{#1}}
\makeatother
\providecommand*\mcitethebibliography{\thebibliography}
\csname @ifundefined\endcsname{endmcitethebibliography}
  {\let\endmcitethebibliography\endthebibliography}{}

\end{document}


\newpage

\setcounter{page}{1}
\makeatletter
\renewcommand{\thepage}{ S\@arabic\c@page }

\setcounter{section}{0}
\makeatletter
\renewcommand{\thesection}{S.\@arabic\c@section }

\setcounter{figure}{0}
\makeatletter
\renewcommand{\thefigure}{S\@arabic\c@figure }  

\setcounter{table}{0}
\makeatletter
\renewcommand{\thetable}{S\@arabic\c@table } 

\setcounter{equation}{0}
\makeatletter
\renewcommand{\theequation}{S.\@arabic\c@equation}

\tableofcontents

\clearpage

\section{\hspace{1mm}Workflow Depicting the Validation of Scattering Models}

\subsection{Quantifying the Errors in R$\mathrm{_g}$}

\begin{figure}[!h]
	\centering
	\includegraphics[width=0.99\textwidth]{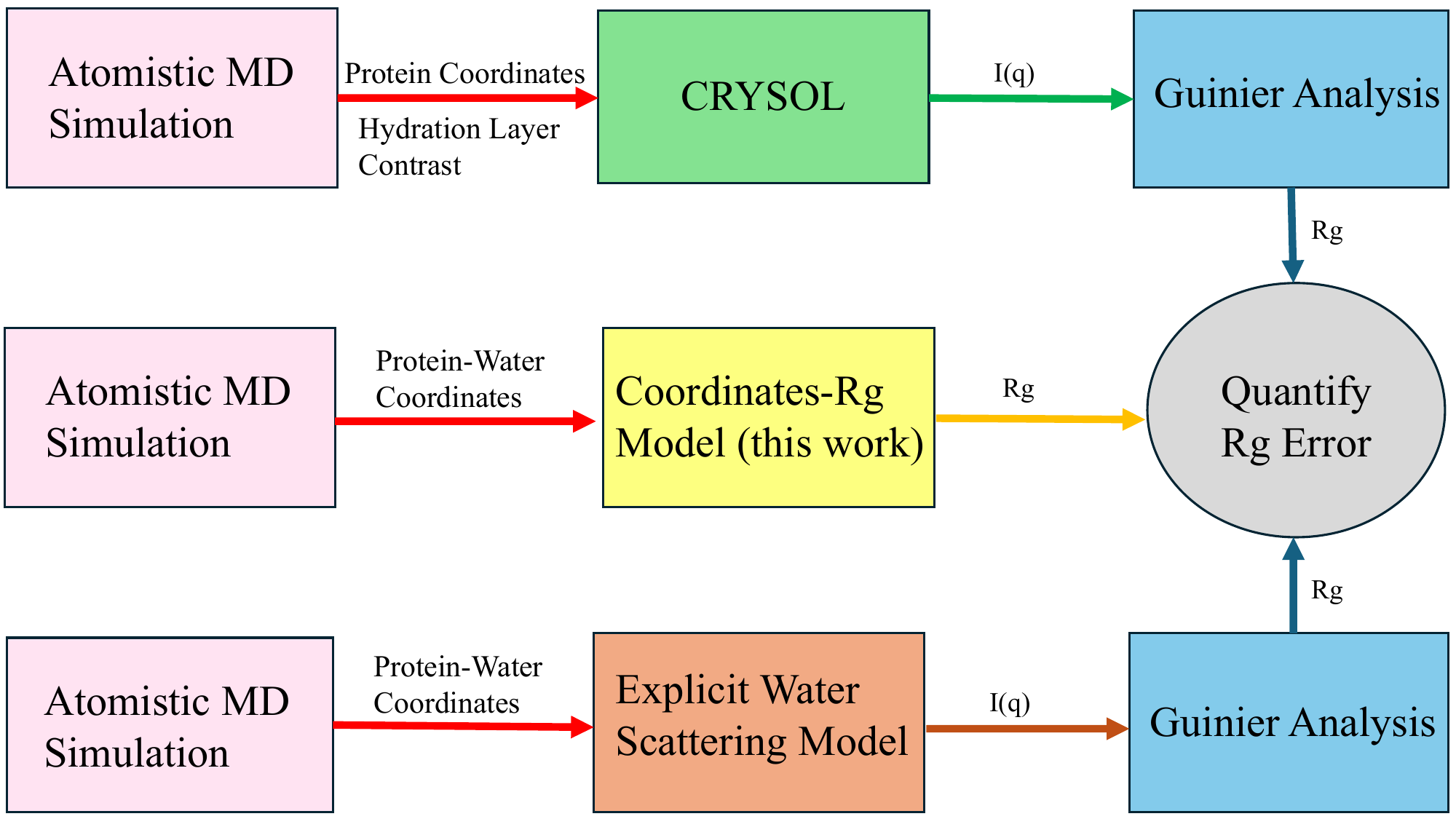}
    \vspace{0.5cm}
	\caption{A flowchart depicting the philosophy of scattering model validation used in this work. An equation for calculating the radius of gyration (R$\mathrm{_g}$) from the coordinates of an atomistic simulation is presented in this work. The R$\mathrm{_g}$ predicted by the CRYSOL model\cite{svergun1995crysol} and the explicit water scattering model\cite{park2009simulated} are compared to the coordinates-based model to quantify the errors in R$\mathrm{_g}$.}
	\label{fig:flowchart_1}
\end{figure}	

\clearpage

\subsection{Quantifying the Over-Fitting from CRYSOL}

\begin{figure}[!h]
	\centering
	\includegraphics[width=0.99\textwidth]{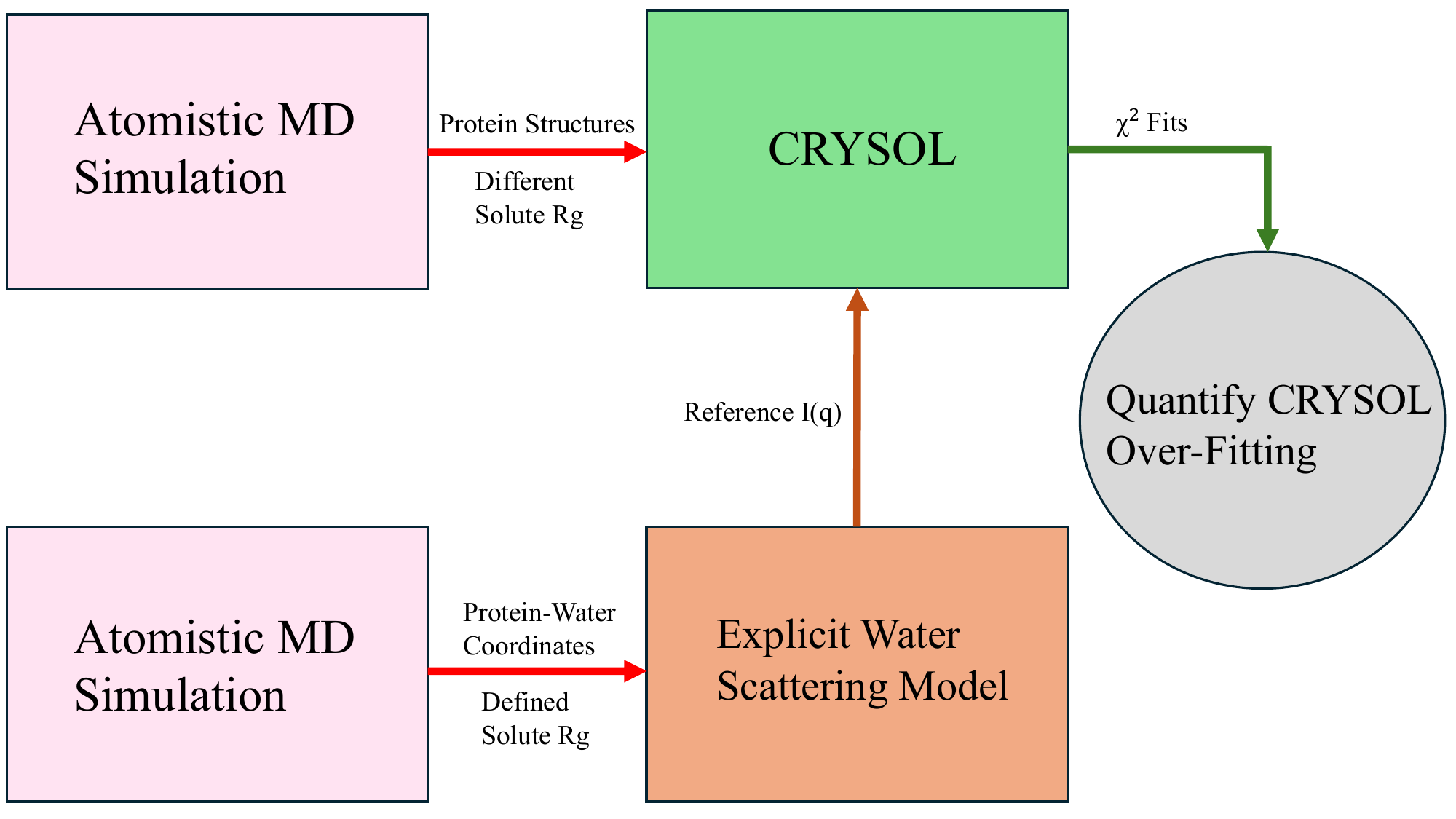}
    \vspace{0.5cm}
	\caption{A flowchart depicting the quantification of over-fitting from the CRYSOL model\cite{svergun1995crysol}. The explicit water scattering model\cite{park2009simulated} with a frozen protein structure is used as a reference. Protein structures with varying solute R$\mathrm{_g}$ are fed into CRYSOL with the I(q) from the explicit water scattering model. The $\chi^2$ fits obtained from CRYSOL are studied as a function of the difference between the R$\mathrm{_g}$ of the solute structure in question and the R$\mathrm{_g}$ of the reference solute structure.}
	\label{fig:flowchart_2}
\end{figure}

\clearpage
 
\section{\hspace{1mm}Simulation Details}
\subsection{Simulations for Error Analysis using Contrast Matching}
For the analysis of errors from the CRYSOL model\cite{svergun1995crysol} and the Explicit Water Scattering Model (EWSM) \cite{park2009simulated} the MD simulations for 20 proteins were performed using the CHARMM36m (C36m)\cite{huang2017charmm36m} force field for proteins and the modified TIP3P waters (mTIP3P)\cite{neria1996simulation}. All of the proteins were capped with ACE and NH2 terminal groups. An appropriate number of Na$^{+}$ or Cl$^{-}$ ions were added to counter balance the charges in the simulation system. The protein-water system containing the solute and 75000 mTIP3P waters were packed into a simulation box using PACKMOL\cite{martinez2009packmol}. The system was then energy minimized for 10000 steps to remove bad contacts. All simulations were carried out at 298.15 K and 1 atm pressure using a Langevin thermostat and a Langevin barostat on NAMD\cite{phillips2020scalable}.  The simulations were run for 4 ns with a 2 fs time step by freezing the protein structure. (Note that one of us (DA) has shown that a 2~fs time-step with a rigid body description of water does violate equipartition between translational and rotational modes of water \cite{Asthagiri:jctc2024a}. The effect is rather subtle, but one that is not expected to alter the main conclusions of this work, which was completed before the aforementioned work.)  

After neglecting the first 2 ns of simulation, structures were saved every 2 ps to obtain a 1000 frames of simulated data. The Lennard-Jones (LJ) and electrostatic potentials were cut-off at 16 $\mathrm{\AA}$ with a switching function for LJ interactions beginning at 15 $\mathrm{\AA}$. Neighbor lists of atom pairs were stored up to a distance of 18 $\mathrm{\AA}$. The long-range contribution to electrostatics was incorporated using the Patricle Mesh Ewald (PME) approach, with a grid spacing of 1 $\mathrm{\AA}$\cite{frenkel2023understanding}. The PDB codes for all 20 proteins and the errors from the scattering models are shown in Table \ref{tab:si_tab_1}. The 20 proteins used in our study were also used to predict the hydration layer around proteins in the HyPred program \cite{virtanen2011modeling}.

\begin{table}[!ht]
		\centering
		\hspace*{-0.5cm}\begin{tabular}{ccccc} 
			\hline\hline
			PDB Code & Number of Residues & Contrast [\%] & $\Delta$R$\mathrm{_g}$ (CRYSOL) [$\mathrm{\AA}$] & $\Delta$R$\mathrm{_g}$ (EWSM) [$\mathrm{\AA}$]   \\  
			\hline
			 	1UBQ &  76  &  19.77    &  2.06 & 0.02 \\
			    6LYZ &  129 &  19.08    &  2.44 & 0.01 \\
                1WLA &  153 &  19.78    &  1.00 & 0.02 \\
                1CRC &  105 &  21.00    &  2.40 & 0.01 \\
                1DF4 &  62  &  15.16    &  2.26 & 0.02 \\
                1DXG &  36  &  19.07    &  0.98 & 0.03 \\
                1F94 &  63  &  17.57    &  2.30 & 0.01 \\
                1HYP &  75  &  15.25    &  2.34 & 0.03 \\
                2L2P &  56  &  16.63    &  1.70 & 0.02 \\
                1TIF &  76  &  18.83    &  2.15 & 0.01 \\
                1UOY &  64  &  19.14    &  2.43 & 0.02 \\
                1US0 & 314  &  23.73    &  1.70 & 0.01 \\
                1VCC & 76   &  18.41    &  2.09 & 0.01 \\
                1YZM & 46   &  16.45    &  1.68 & 0.03 \\
                2DOB & 82   &  19.93    &  1.48 & 0.01 \\
                2PTN & 230  &  19.79    &  1.82 & 0.02 \\
                2ZQE & 80   &  19.35    &  2.36 & 0.01 \\
                3G19 & 84   &  18.52    &  1.80 & 0.01 \\
                3HGL & 77   &  19.65    &  2.40 & 0.02 \\
                3LE4 & 55   &  15.10    &  2.55 & 0.02 \\
			\hline\hline
		\end{tabular}
		\caption{Details of the 20 proteins used in the analysis of the CRYSOL scattering model and the explicit water scattering model (EWSM). The errors from CRYSOL remain significant irrespective of the contrast in the hydration layer or the size of the system.}
		\label{tab:si_tab_1}
	\end{table}	

\subsection{Simulations for the $\mathrm{\chi^2}$ Fit Analysis using CRYSOL}
For the computation of $\mathrm{\chi^2}$ as a function of $\Delta$R$\mathrm{_g}$ (solute), 4 protein-water systems were simulated in the NVT ensemble. The simulations were carried out at high temperature (1000K) using a 1 fs time step for 2 ns. The average volume of the system from the previous simulation was used to define the simulation cell dimensions. 
 Thermal fluctuations of the protein at high temperatures creates an R$\mathrm{_g}$ variation in the solute. 8 structures with R$\mathrm{_g}$ increasing in 0.5 $\mathrm{\AA}$ intervals were selected from the structures obtained from each of the 4 simulations. The reference solute structure was chosen with an R$\mathrm{_g}$ 2.5 $\mathrm{\AA}$ higher than the most compact selected structure and 1 $\mathrm{\AA}$ lower than the most extended selected structure. The reference solute-water system was once again simulated by freezing the solute at 298.15 K and 1 atm (NPT ensemble) using the procedure described above. Cutoffs for the various interactions were defined as described previously.

\section{\hspace{1mm}Contrast Estimation}

\subsection{Determination of the Protein Surface}
The contrast in the 3 $\mathrm{\AA}$ hydration shell around the solute is an important parameter that can be input into CRYSOL. To calculate the contrast in the system, the excluded volume of the solute first needs to be defined. We follow Virtanen et.~al.\cite{virtanen2011modeling}~and define around 300 atom types to define a detailed surface for the protein. Each atom of each of the 20 amino acids constitutes a separate atom type. This is more detailed than previous studies where the protein is divided into 5 atom groups (C, N, O, S, H)\cite{virtanen2010modeling}. We simulated linear chains of each of the 20 amino acid residues in 34000 mTIP3P waters. The simulations were run at 298.15 K and 1 atm pressure for 4 ns with a 2 fs time step. Trajectories were stored every 2 ps in the last 2 ns after neglecting the first 2 ns. The closest distance of any water atom in the 1000 frames of simulated data to each of the atoms of the amino acid residues were recorded. This is treated as the surface radii of that particular atom type. The surface of the protein is then defined based on its amino acid sequence using the surface radii of the various atom types. 

\subsection{Calculation of the Hydration Layer Contrast}
Once the protein surface was defined, the hydration layer was defined as the lowest volume that encapsulates all points that are less than 3 $\mathrm{\AA}$ ($>$ 0 $\mathrm{\AA}$) away from the protein surface. The contrast was then calculated using a reference bulk water simulation. The number of waters in the hydration shell surrounding the protein averaged over 1000 frames of the simulation ($\bigl< \mathrm{N_{w}^{p}} \bigl>$) and the number of waters in the same volume of the bulk water simulation ($\bigl< \mathrm{N_{w}^{b}} \bigl>$) were used to calculate the contrast in the hydration layer as shown in Eq.~\ref{eq:eq_contr}

\begin{equation}
    \delta \rho\mathrm{_h} = \biggl( \frac{\bigl< \mathrm{N_{w}^{p}} \bigl> - \bigl< \mathrm{N_{w}^{b}} \bigl>}{\bigl< \mathrm{N_{w}^{b}} \bigl>} \biggl) \rho\mathrm{_{b}}
    \label{eq:eq_contr}
\end{equation}

$\mathrm{\rho_{b}}$ is the electron density of the bulk solvent, which is 0.334 $\mathrm{e/\AA^3}$ for water at 298.15 K and 1 atm pressure. For modified TIP3P waters the bulk electron density is slightly higher at 0.3389 $\mathrm{e/\AA^3}$, which is the value we use in our calculations. The contrast calculated for each of the 20 proteins is shown in Table \ref{tab:si_tab_1}. The contrast so calculated ($\mathrm{\delta \rho_h}$) and the bulk density of the buffer ($\rho\mathrm{_{b}}$) were input into CRYSOL to analyze the errors from the model.  

\clearpage

\section{\hspace{1mm}Scattering and R$\mathrm{_g}$ Computation}

\subsection{Scattering Computations for Error Analysis using Contrast Matching}

The explicit water scattering model (EWSM)\cite{park2009simulated} is shown in Eq.~\ref{eq:si_expl_sca}, where A(\textbf{q}) is the amplitude from the protein-water simulation and B(\textbf{q}) is the amplitude from the bulk water simulation.

\vspace{-0.75cm}
\begin{equation}
        \Delta I(q) = \bigg \langle \big | \langle A(\textbf{q}) \rangle - \langle B(\textbf{q}) \rangle \big | ^2 + 
        \big[ \langle |A(\textbf{q})|^2 \rangle  -
         \big | \langle A(\textbf{q}) \rangle \big |^2 \big] -  
        \big[ \langle |B(\textbf{q})|^2 \rangle  - \big | \langle B(\textbf{q}) \rangle \big |^2 \big] \bigg \rangle_{\mathrm{\Omega}}
        \label{eq:si_expl_sca}
    \end{equation}

For the calculation of $\Delta$R$\mathrm{_g^{EWSM}}$ by contrast matching, the equations to calculate A(\textbf{q}) and B(\textbf{q})\cite{chen2014validating} are shown in Eq.~\ref{eq:con_match_a_b}. 

\vspace{-0.3cm}
\begin{align}
        & A(\textbf{q}) = \sum_{j = 1}^{N_{A}} n_e^{A} e^{-i\textbf{q}.r_{j}}, \nonumber \\
        & B(\textbf{q}) = \sum_{j = 1}^{N_{B}} n_e^{B} e^{-i\textbf{q}.r_{j}},
        \label{eq:con_match_a_b}
\end{align}

Here the Fourier transform of the atomic coordinates are weighted by the electron numbers n$\mathrm{_e^{A}}$ and n$\mathrm{_e^{B}}$ to maintain consistency with the R$\mathrm{_g}$ calculation from coordinates. The background subtracted intensities from SAXS measurements are orientationally averaged due to the presence of a very high number of solutes which sample all orientations in solution\cite{svergun2013small}. The intensities calculated from simulation data also need to be orientationally averaged to match the experimental procedure. We perform an orientational average using the method of spirals\cite{ponti1999simulation} and 1500 direction vectors. Using 1500 direction vectors was found to be sufficient to obtain averages within 1 \% error\cite{park2009simulated}. The waters in both the protein-water and bulk water simulations were chosen if they were within $3\, \mathrm{\AA}$ of the protein surface to maintain consistency with CRYSOL\cite{manalastas2021atsas}.

The R$\mathrm{_g}$ from the scattering curve was obtained using Guinier analysis\cite{cantor1980biophysical}. In our calculation, we used 10 data points in the q range of [0.001, 0.01] $\mathrm{\AA^{-1}}$ to obtain R$\mathrm{_g}$ from the slope of the intensities as shown in Eq.~\ref{eq:guinier}. 

\vspace{-0.3cm}
\begin{eqnarray}
     \mathrm{ln}\bigg[\frac{I(q)}{I(0)}\bigg] \approx -\frac{q^2R_g^2}{3}
\label{eq:guinier}
\end{eqnarray}

\subsection{Scattering Computations for the $\mathrm{\chi^2}$ Fit Analysis using CRYSOL}

For the calculation of $\mathrm{\chi^2}$ fits from CRYSOL, EWSM is used as the reference. In this case the role of EWSM is to mimic experimental data. Hence we select all waters that are less than $7\, \mathrm{\AA}$ away from any solute atom in accordance with Ref.~\citenum{park2009simulated}. The amplitude from the protein-water and bulk water simulations are now calculated using Eq.~\ref{eq:expl_sca_elec_dens} which incorporates the electron density clouds (form factors) of atoms (f$\mathrm{_j}$(q)).    

\vspace{-0.3cm}
\begin{align}
        & A(\textbf{q}) = \sum_{j = 1}^{N_{A}} f_j(q) e^{-i\textbf{q}.r_{j}}, \nonumber \\
        & B(\textbf{q}) = \sum_{j = 1}^{N_{B}} f_j(q) e^{-i\textbf{q}.r_{j}}
        \label{eq:expl_sca_elec_dens}
\end{align}

To account for the polarizability of water, Sorenson et.~al\cite{sorenson2000can}.~proposed a correction to f$\mathrm{_j}$(q) for water oxygens and hydrogens which is implemented in our work. The amplitudes A and B so calculated were then input into Eq.~\ref{eq:si_expl_sca} to calculate the background subtracted intensities. The background subtracted intensities were calculated for 101 $q$ values in the range of [0.0, 0.5] $\mathrm{\AA^{-1}}$. The orientational averages were once again computed from the spiral method using 1500 direction vectors. We followed Park et.~al\cite{park2009simulated}.~and divided the 100 simulation frames into groups of 10 to obtain the error bars from block averaging.

\subsection{Computation of R$\mathrm{_g}$ from Simulation Coordinates}

In this work we have developed an equation for the radius of gyration from the coordinates of an explicit water simulation (R$\mathrm{_g^{coor}}$). This is used as a benchmark to validate the R$\mathrm{_g}$ prediction from scattering models. In this section we show the steps involved in the development of Eq.~4 (main text). As we walk through the steps we appeal to the intuition of the reader. We begin with the equation for the radius of gyration for a system of N components from polymer theory (Eq.~\ref{eq:rg_polymer_theory})\cite{ivankov2009coupling}.

\begin{eqnarray}
        R_{g}^{2} =   \frac{\sum\limits_{i=1}^{N} m_i|\mathbf{r_i} - \mathbf{r_{com}}|^2}{\sum\limits_{i=1}^{N}m_i} 
        \label{eq:rg_polymer_theory}
\end{eqnarray}

Since we are comparing the R$\mathrm{_g}$ from coordinates to the predictions from a scattering model, we recognize that X-rays interact with the electrons of the system and not the nucleus. Hence we replace the masses of the atoms (m$\mathrm{_i}$) in Eq.~\ref{eq:rg_polymer_theory} with their electron numbers (n$\mathrm{_i^{e}}$). Which gets us to Eq.~\ref{eq:rg_polymer_electron}.

\begin{eqnarray}
        R_{g}^{2} =   \frac{\sum\limits_{i=1}^{N} n_i^e|\mathbf{r_i} - \mathbf{r_{com}}|^2}{\sum\limits_{i=1}^{N}n_i^e}
        \label{eq:rg_polymer_electron}
\end{eqnarray}

In background subtraction, the I(q) of bulk waters is subtracted from the I(q) of the protein-water solution. Hence we need to separate the number of atoms (N) in Eq.~\ref{eq:rg_polymer_electron} into the atoms in the protein solution (N$\mathrm{_A}$) and atoms in the buffer solution (N$\mathrm{_B}$). The atoms from the bulk water simulation (N$\mathrm{_B}$) have negative weights to maintain consistency with the background subtraction procedure. With these modifications to Eq.~\ref{eq:rg_polymer_electron} we obtain Eq.~\ref{eq:rg_polymer_subtraction}

\begin{eqnarray}
        R_{g}^{2} =  \frac{\sum\limits_{N_A} n_A^e|\mathbf{r_A} - \mathbf{r_{com}}|^2 - \sum\limits_{N_B} n_B^e|\mathbf{r_B} - \mathbf{r_{com}}|^2}{\sum\limits_{N_A}n_A^e - \sum\limits_{N_B}n_B^e} 
        \label{eq:rg_polymer_subtraction}
\end{eqnarray}

In SAXS measurements, the amplitudes from different conformations of proteins add up and the intensities captured by the receptor (square of the magnitudes of the amplitudes) are a reflection of the properties of the ensemble of structures. To mirror this, the information from the different frames of a simulation need to be superimposed so that the R$\mathrm{_g}$ obtained is a property of the simulation system as a whole. We take this fact into account by summing over the information from all frames of the simulation (N$\mathrm{_F}$). Which leaves us with Eq.~\ref{eq:si_rg_coor}. The number of atoms in the volume occupied by the protein and the hydration layer (N$\mathrm{_A}$) and the number of bulk water atoms in the same volume (N$\mathrm{_B}$) are functions of the simulation frame in consideration.

\begin{eqnarray}
        R_{g}^{coor} =  \sqrt{\frac{\sum\limits_{N_F} \sum\limits_{N_A} n_{A}^{e} |\mathbf{r_{A}} - \mathbf{r_{com}}|^2 - \sum\limits_{N_F} \sum\limits_{N_B} n_{B}^{e}|\mathbf{r_{B}} - \mathbf{r_{com}}|^2}{\sum\limits_{N_F} \sum\limits_{N_A} n_{A}^{e} - \sum\limits_{N_F} \sum\limits_{N_B} n_{B}^{e}}} 
        \label{eq:si_rg_coor}
\end{eqnarray}

Similarly the equation for the center of mass transforms from Eq.~\ref{eq:com_polymer_theory}\cite{ivankov2009coupling} to Eq.~\ref{eq:si_coor_com}.

\begin{eqnarray}
         \mathbf{r_{com}} = \frac{\sum\limits_{i=1}^N m_i \mathbf{r_i}}{\sum\limits_{i=1}^N m_i}
    \label{eq:com_polymer_theory}
\end{eqnarray}

\begin{eqnarray}
         \mathbf{r_{com}} = \frac{\sum\limits_{N_F} \sum\limits_{N_A} n_{A}^{e} \mathbf{r_{A}} - \sum\limits_{N_F} \sum\limits_{N_B} n_{B}^{e}\mathbf{r_{B}}}{\sum\limits_{N_F} \sum\limits_{N_A} n_{A}^{e} - \sum\limits_{N_F} \sum\limits_{N_B} n_{B}^{e}}
\label{eq:si_coor_com}
\end{eqnarray}

Eq.~\ref{eq:si_rg_coor} is general and can be applied for any thickness of the hydration shell. Just as in the explicit water scattering model, the division of the protein-hydration layer complex into an excluded volume and hydration layer is also not necessary. The equation is also applicable for a control volume defined on the basis of the distance from the center of protein atoms instead of the protein surface. In this work to maintain consistency with CRYSOL, we define the hydration layer to be $3\, \mathrm{\AA}$ away from the surface of the protein in all of our R$\mathrm{_g^{coor}}$ calculations. The R$\mathrm{_g^{coor}}$ calculation from Eq.~\ref{eq:si_rg_coor} and the calculation of intensities from the explicit water scattering model (Eq.~\ref{eq:si_expl_sca}) were performed using in-house PYTHON scripts built atop the MDTraj library\cite{McGibbon2015MDTraj}.   

\clearpage

\section{\hspace{1mm}Validation of the Scattering Code}

To ensure that EWSM is being implemented correctly, we repeated the calculation in Ref.~\citenum{park2009simulated} for lysozyme (PDB code: 6LYZ). Lysozyme was modeled using the CHARMM22 (C22)\cite{mackerell1998all} force field and simulated in 34000 TIP3P\cite{jorgensen1983comparison} waters on NAMD\cite{phillips2020scalable}. The solute was capped with N and C terminal groups using psfgen. The simulation was carried out at 277.15 K and 1 atm pressure using a 2 fs time step for 0.4 ns. Structures were saved every 2 ps after discarding the first 0.2 ns to obtain 100 frames of simulated data. The same procedure was used to simulate 34000 TIP3P waters and obtain 100 frames of data for the bulk solvent. Lennard-Jones (LJ) and electrostatic interactions were cut off at $16\, \mathrm{\AA}$. LJ was gradually turned off by switching the interactions at $15\, \mathrm{\AA}$. For long-range electrostatics we used PME with a grid spacing of $1\, \mathrm{\AA}$. Waters were chosen if they were less than $7\, \mathrm{\AA}$ away from any solute atom. This envelope is large enough to account for all non bulk-like behavior of waters around the solute \cite{chen2014validating}.

\begin{figure}[!h]
	\centering
    \vspace{-1.0cm}
	\includegraphics[width=0.99\textwidth]{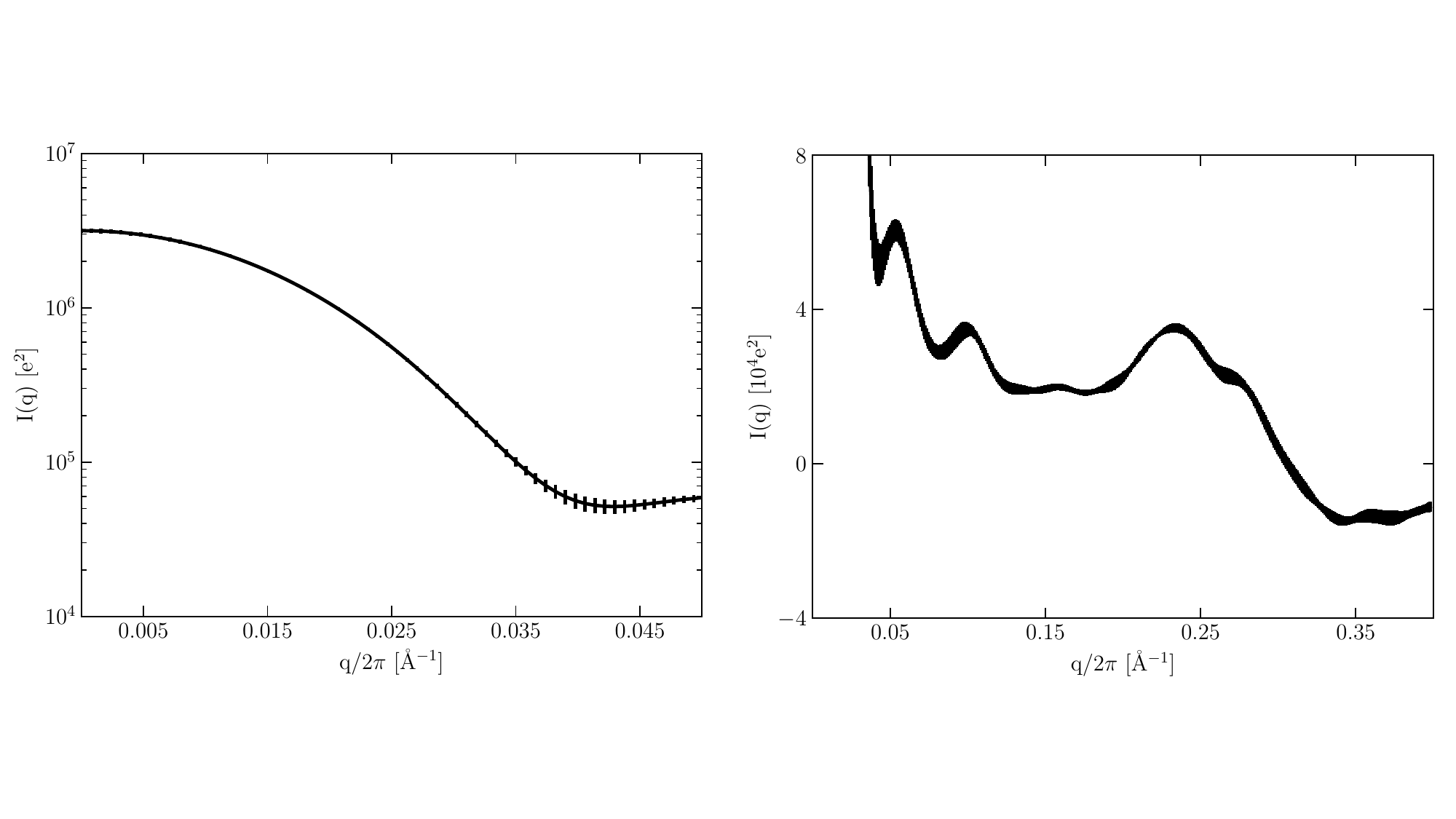}
    \vspace{-1.3cm}
	\caption{The background subtracted intensities for lysozyme are plotted against the magnitude of the momentum transfer vector (\textbf{q/$\mathrm{2\pi}$}). The intensities at low angle (q/$2\pi$ $<$ 0.05 $\mathrm{\AA^{-1}}$) are best represented on the log scale (left). The intensities at higher angles (q/$2\pi$ $>$ 0.05 $\mathrm{\AA^{-1}}$) have peaks and valleys that are best captured in the normal cartesian representation (right). The results are in excellent quantitative agreement with Park et.~al\cite{park2009simulated}.}
	\label{fig:lyz_sca_val}
\end{figure}

The scattering model shown in Eq.~\ref{eq:si_expl_sca} was implemented using PYTHON scripts and the 100 frames were grouped into 10 blocks to estimate the error bars. The background subtracted intensities for lysozyme are shown in Fig.~\ref{fig:lyz_sca_val}. The results are in excellent quantitative agreement with the predictions made by Park et.~al\cite{park2009simulated}. The validated scattering code was then used to calculate the EWSM intensities for all the systems described in this study.

\clearpage

\section{\hspace{1mm}The Effect of Thermal Fluctuations on R$\mathrm{_g}$}

In this section we enquire if the thermal fluctuations of the solute causes a significant change in the R$\mathrm{_g}$ computed from Guinier analysis.

\subsection{Details for the Thermally Fluctuating Protein Simulations}

 We took 4 proteins (PDB codes: 1DXG, 1F94, 1YZM, and 3LE4) and simulated them in explicit water by allowing for thermal fluctuations. The simulations were run at 298.15 K and 1 atm pressure for 30 ns. The first 10 ns were neglected and the trajectory file was updated every 2 ps. Out of the 10000 frames of simulated data, 1000 frames were chosen to emulate a Gaussian distribution in the R$\mathrm{_g}$ (solute) space. A protein structure with R$\mathrm{_g}$ at the center of the normal distribution (maximum probability) was chosen for the fixed solute scattering analysis. The simulation for the fixed solute in explicit water was performed as described previously and 1000 frames of simulation data were obtained. Scattering computations on the system were performed as described before. Error bars were calculated by block averaging using 5 blocks. 

\subsection{Scattering Computation for Thermally Fluctuating Proteins}

The equation to calculate the scattering curve of a thermally fluctuating protein in explicit water is the same as the one for a frozen solute (Eq.~\ref{eq:si_expl_sca}). However, the volume of the chosen envelope (control volume) varies \cite{chen2014validating, knight2015waxsis}. To study fluctuating proteins, the volume of the envelope must be large enough to contain the density fluctuations in water (around the solute) for every configuration. Knight and Hub\cite{knight2015waxsis} define the volume as an icosphere with 5120 vertices. In this work we divided the volume in $\mathrm{\phi-\psi}$ space by using 10370 direction vectors. The 10370 direction vectors were moved from the center of the simulation box until they were $7\, \mathrm{\AA}$ away from all solute atoms in all simulation frames. A water molecule was either chosen or neglected for analysis based on the r, $\mathrm{\theta}$, $\mathrm{\phi}$ coordinates of its atoms. If any water atom was closer to the center of the box than the maximum radii in the quadrant [r$\mathrm{_1}$, $\mathrm{\theta_1}$, $\mathrm{\phi_1}$], [r$\mathrm{_2}$, $\mathrm{\theta_1}$, $\mathrm{\phi_2}$], [r$\mathrm{_3}$, $\mathrm{\theta_2}$, $\mathrm{\phi_1}$], and [r$\mathrm{_4}$, $\mathrm{\theta_2}$, $\mathrm{\phi_2}$] defined by the direction vectors, it was chosen for analysis. A similar procedure was used to carve out an envelope in the bulk water simulation. The definition of the envelope and the calculation of the scattering from the solute in every frame are the only two differences between the scattering computation for a fluctuating solute and a frozen solute \cite{hub2018interpreting}. 

\subsection{Comparison of Scattering Methods}

The contrast in the $3\, \mathrm{\AA}$ hydration shell is compared for the fluctuating and frozen solutes in Table \ref{tab:si_tab_2}. For the contrast calculation of a thermally fluctuating protein, the hydration layer is dynamic and needs to be redefined every frame. The rest of the analysis remains the same. The contrast for all of the 4 proteins from the two methods are sufficiently close. There are regions around the solute that are quite favorable to the solvent, as the solute tumbles through the solution, the solvent molecules remain bound to the protein and move along with the solute. This is the reason the contrast calculated from the two methods remains fairly similar.

\begin{table}[!ht]
		\centering
		\hspace*{-0.5cm}\begin{tabular}{ccccc} 
			\hline\hline
			PDB Code & Contrast (frozen) [\%] & Contrast (fluctuating) [\%] & R$\mathrm{_g}$ (frozen) [$\mathrm{\AA}$] & R$\mathrm{_g}$ (fluctuating) [$\mathrm{\AA}$]   \\  
			\hline
			 1DXG &  $17.48 \pm 0.33$  &  17.48    &  $11.95 \pm 0.14$ & 12.16 \\
                          1F94 &  $16.80 \pm 0.09$  &  16.29    &  $11.93 \pm 0.20$ & 11.48 \\
                        1YZM &  $15.82 \pm 0.24$  &  15.87    &  $12.16 \pm 0.12$ & 12.29 \\
                         3LE4 &  $15.15 \pm 0.15$  &  15.41    &  $12.03 \pm 0.29$ & 11.48 \\
			\hline\hline
		\end{tabular}
		\caption{Comparison of the hydration layer contrast and the radius of gyration (R$\mathrm{_g}$) from Guinier analysis for the frozen and fluctuating solute system. Both methods yield similar properties on the macroscale.}
		\label{tab:si_tab_2}
	\end{table}	

\begin{figure}[!h]
	\centering
    \vspace{-0.5cm}
	\includegraphics[width=0.99\textwidth]{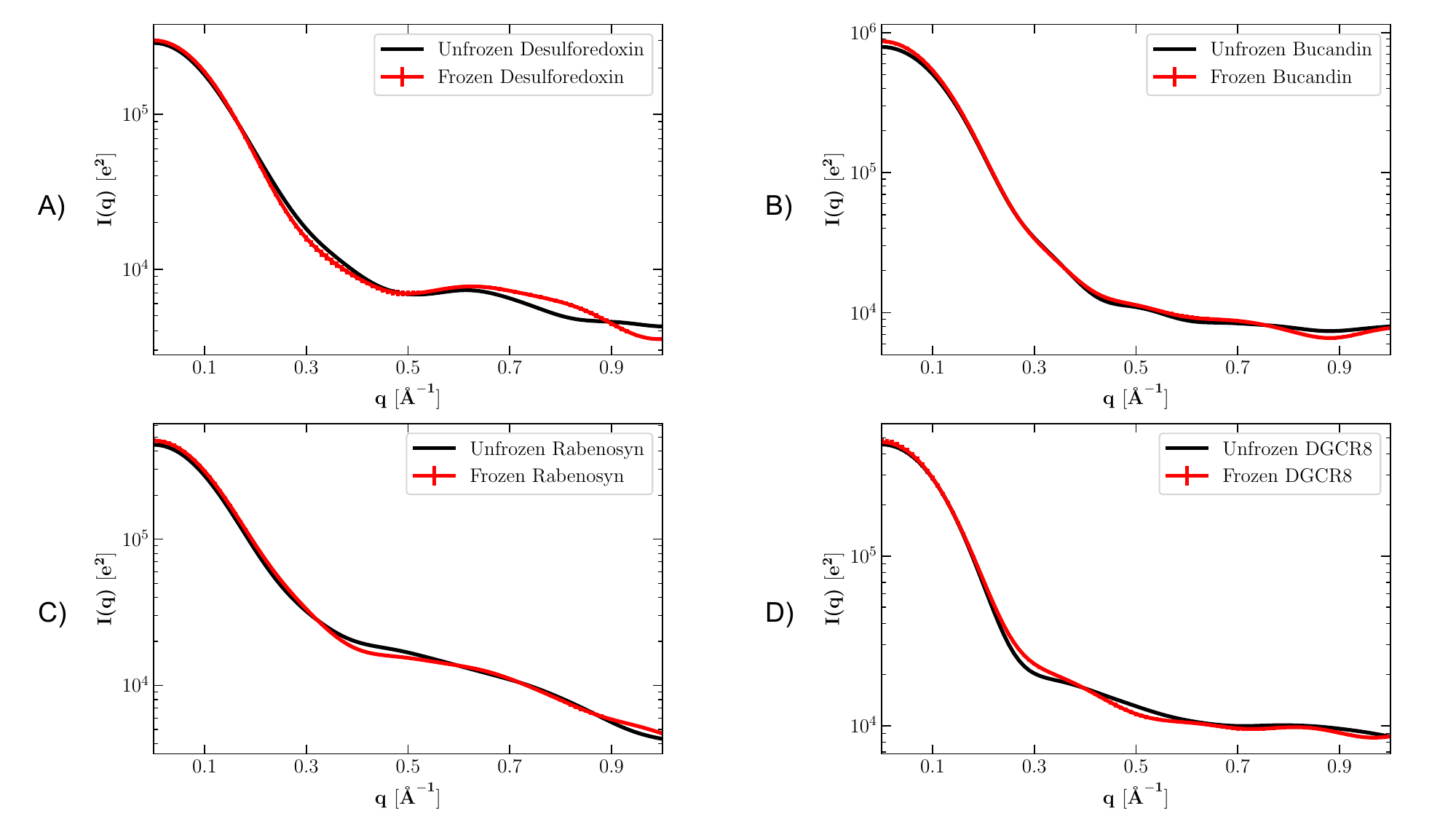}
	\caption{Comparison of the scattering curves from the two methods. Namely a frozen solute in explicit water (red) and a thermally fluctuating solute in explicit water (black). The error bars in the frozen solute scattering analysis are obtained by block averaging. The four proteins studied are desulforedoxin (1DXG) in (A), bucandin (1F94) in (B), rabenosyn (1YZM) in (C), and DGCR8 (3LE4) in (D).}
    \label{fig:si_fig_2}
\end{figure}

Scattering curve from the two methods are compared in Fig.~\ref{fig:si_fig_2}. The two curves frequently overlap in the Guinier region. The R$\mathrm{_g}$ from Guinier analysis is also reported in Table \ref{tab:si_tab_2}, and is sufficiently close. This is due to the fact that the relative distance between the solute atom and the solvent molecules is preserved on a macroscale even as the protein tumbles in solution, causing little to no effect on the system R$\mathrm{_g}$. 

The primary reason for the minor deviations in R$\mathrm{_g}$ and contrast is due to the comparison between two different ensembles. In our comparison of R$\mathrm{_g^{CRYSOL}}$, R$\mathrm{_g^{EWSM}}$, and R$\mathrm{_g^{coor}}$, all of the calculations pertained to the same ensemble, even though CRYSOL is a low information representation of the explicit solvent ensemble \cite{park2009simulated}. This allowed us to negate the variations that naturally arise during ensemble generation and isolate the errors from the scattering models. In our comparison of fluctuating and frozen proteins, the ensemble is necessarily different because they relate to different simulations. Hence it is not possible to isolate the differences in the scattering methods. 

The average density of bulk waters over a 1000 frames is never identical between two independent simulations. There are always minor fluctuations in the average density. Chen et.~al\cite{chen2014validating}.~tracked these variations to tangible differences in intensities (I(q)) at low scattering angles. This probably explains the minor differences between the methods. Despite these complexities, the contrast and the R$\mathrm{_g}$ from the two methods are sufficiently close and we conclude that the thermal fluctuations alone does not impact the R$\mathrm{_g}$ of the system.

\clearpage


\providecommand{\latin}[1]{#1}
\makeatletter
\providecommand{\doi}
  {\begingroup\let\do\@makeother\dospecials
  \catcode`\{=1 \catcode`\}=2 \doi@aux}
\providecommand{\doi@aux}[1]{\endgroup\texttt{#1}}
\makeatother
\providecommand*\mcitethebibliography{\thebibliography}
\csname @ifundefined\endcsname{endmcitethebibliography}
  {\let\endmcitethebibliography\endthebibliography}{}